\documentclass[aps,prb,twocolumn,epsfig,showpacs]{revtex4}
\usepackage{graphicx}
\begin{document}

\title{Double dynamical regime of confined water}
\author{P.~Gallo and M.~Rovere}
\address{Dipartimento di Fisica, 
Universit\`a ``Roma Tre'', \\ Istituto Nazionale per la Fisica della Materia,
 Unit\`a di Ricerca Roma Tre\\
Via della Vasca Navale 84, 00146 Roma, Italy.}

\begin{abstract}
The Van Hove self correlation function of water confined in a
silica pore is calculated from Molecular Dynamics trajectories
upon supercooling. At long time in the $\alpha$ relaxation region
we found that the behaviour of the real space time dependent correlators 
can be decomposed in a very slow, almost frozen, dynamics due to
the bound water close to the substrate and a faster dynamics of
the free water which resides far from the confining surface.
For free water we confirm the evidences of an approach to a 
crossover mode coupling transition, previously found in 
Q space. In the short time region we found that
the two dynamical regimes are overimposed and cannot be
distinguished. This shows that the interplay between the
slower and the faster dynamics emerges in going from early times to
the $\alpha$ relaxation region, where  
a layer analysis of the dynamical properties can be performed.
\end{abstract}

\pacs{61.20.Ja, 61.20.-p, 61.25.-f}

\maketitle

\section{ Introduction}

The behavior of water below its freezing point upon supercooling it
is not easily investigated by experimentalists since nucleation
processes drive the liquid toward its crystalline phase~\cite{pablo}. 
This phenomenon prevents the
observation of the transition of the supercooled liquid to the
glassy phase. 
Molecular dynamics simulations~\cite{gallo-prl} predict 
a dynamical transition in the supercooled region 
well described by Mode Coupling Theory (MCT)~\cite{goetze}. 
In the idealized version of MCT the liquid undergoes
a structural arrest at a temperature $T_C$. This 
temperature in real liquids coincides with the cross over temperature 
at which the structural relaxations of the supercooled
liquid are frozen. In most liquids hopping processes start occurring close to
and below $T_C$ and these processes 
ensure ergodicity to the system in this region.
When also hopping
is frozen the thermodynamic glass transition takes place. Ideal predictions
of MCT are well tested on approaching $T_C$ in the region where hopping 
processes are negligible. 

Increasing interest in the study of water when confined arises mainly
because modifications of the behavior of this liquid with respect
to the bulk phase are closely connected to technological and biophysical 
problems. In particular water seems to be more easily supercooled
when confined. This might open an experimental window in a region
of the phase diagram that is experimentally not accessible for the bulk. 

Both simulations and theoretical modelizations of water confined between
two parallel hydrophobic walls predict new scenarios for thermodynamics and
phase behavior~\cite{pablojcp,genejpc}.
Recent experiments on water confined between two slits of mica found 
that hydrophilic confinement seems to primarily suppress the hydrogen bond 
network associated with the freezing~\cite{raviv}. Similar results
have been found in the case of a two
dimensional confinement in vermiculite
clay~\cite{bergman} and in molecular dynamics of water 
in spherical cavities~\cite{geiger}.
A slow relaxation was found for water close to the surface of
proteins~\cite{doster,bizzarri}.
It has also been observed more specifically in the region around
freezing that in water confined in different environments 
the molecules close to the substrate behave differently with respect
to the molecules in the middle of the pores. 
The term of bound water as distinct from free water has been
introduced in order to distinguish the portion of
water which resides in layers close to the surface 
and does not show a real freezing transition, from the
water far from the surface which behaves more similarly to
the bulk water~\cite{morishige}.

Recent neutron diffraction experimental studies on 
water confined in the hydrophilic nanopores  
of Vycor glass have also evidenced
a severe distortion of the hydrogen bond network~\cite{mar12}.
The slow, $\alpha$, relaxation typical of glass formers
has been clearly observed on water confined in Vycor upon supercooling
by means of both a very refined spin-echo technique~\cite{chen-prl}
and a quasi elastic neutron scattering experiment~\cite{zanotti99}.
Evidence of low frequency scattering excess typical of strong
glass formers has been also observed on the same system
for low hydration levels of the pore~\cite{fede}.

We have conducted 
computer simulation work on a model for water
confined in Vycor~\cite{noi-prl,noi-jcp} where  
we found that microscopic forces
due to an hydrophilic surface can induce a layering effect
in water with the formation of a double layer structure
close to the surface. 

In our previous works on dynamics~\cite{noi-prl,noi-jcp}  
we concentrated mainly
on the study of the molecules which reside on the average 
far from the solid surface identified as the free water,
in particular we studied the intermediate scattering function
in the $\alpha$-relaxation region.
MCT appears to work well in its idealized form for free water.
We were able to estimate the cross-over temperatures and other
MCT parameters
for the full hydration~\cite{noi-prl} and for the 
half hydration~\cite{noi-jcp} cases. 

In this paper we present  results obtained from
Molecular Dynamics (MD) simulation at full hydration
which show a clear evidence that our way of separating 
water molecules in two sets, namely the very slow bound water 
that resides in the first two layers close to the substrate and 
the faster free water in the inner
part of the pore is unique for our system, but only in the time window
of the $\alpha$-relaxation.
This effect is in fact clearly visible in the
Van Hove self correlation function  
of the confined water $G_S(r,t)$, 
presented here for the full hydration of the pore. 

In the next section after introducing  the Van Hove self correlation function
we show its behavior in supercooled bulk water
for a comparison with the confined case. In Sec. III we 
give some details about our simulation of confined water
and show the layering effect which takes place in our system.
In Sec. IV we present and discuss the results obtained
for the Van Hove self correlation functions of the
confined water. The final section is devoted to the conclusions. 

\begin{figure}
\includegraphics[width=8cm]{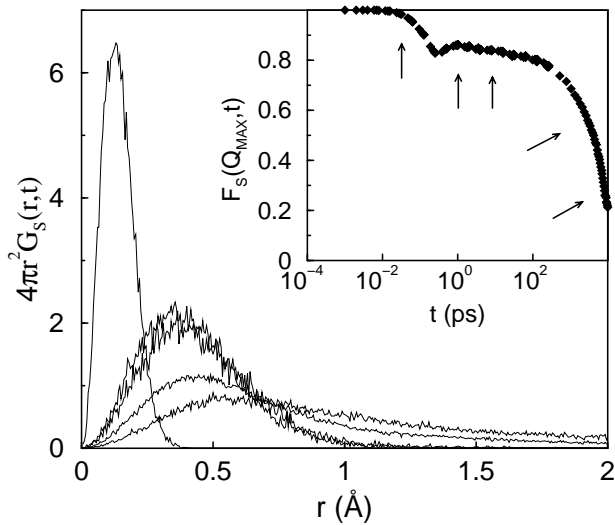}
\caption{In the main frame the angular average of the oxygens
Van Hove self correlation function calculated from MD trajectories
of SPC/E bulk water is displayed for T=190K and density $1$ $g/cm^3$.
Curves from the left correspond to $t=0.032, 1.024, 8.2, 2620$ and
$13107$ $ps$. In the inset the corresponding intermediate scattering function 
at the peak of the structure factor. 
The arrows indicate the times at which the VHSCF is evaluated.
Arrows on the left correspond to curves on the left.}
\protect\label{fig:1}
\end{figure} 
\begin{figure}
\includegraphics[width=8cm]{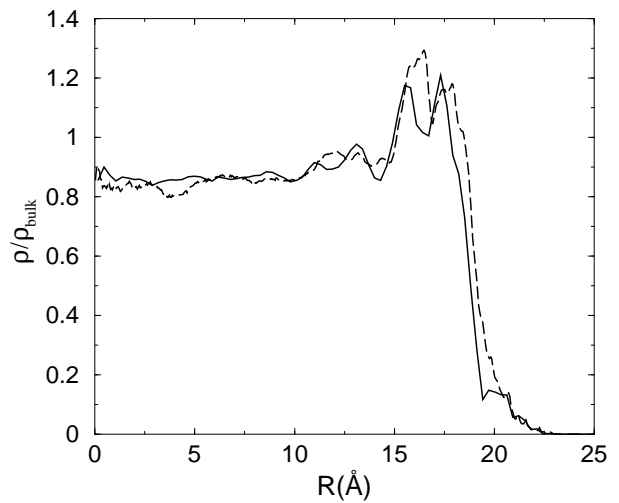}
\caption{Density profiles for confined water normalized to the
bulk value for $T=298$~K  (continuous line) and $T=210$~K long dashed line.
The surface of the Vycor glass is located around $20$ \AA.}
\protect\label{fig:3}
\end{figure} 

\section{The Van Hove self correlation function of bulk water}
    
The Van Hove self correlation function (VHSCF) is defined for a system
of $N$ particles or molecules as 
\begin{equation}
G_S (r,t) = \frac{1}{N} \left< \sum_{i=1}^{N} \delta \left[
{\mathbf{r}} + {\mathbf{r}}_i(0) - {\mathbf{r}}_i(t) \right] 
\right> \label{grts}
\end{equation} 
This function describes the correlation in the
positions of the same atom at different times and more precisely 
$4 \pi r^2 G_S (r,t) d{\mathbf{r}}$ is  
the probability of finding
a particle at distance $r$ after a time $t$ if the same particle
was in the origin $\mathbf{r}=0$ at the initial time $t=0$. 
The Fourier transform of (\ref{grts}) is the  
incoherent or self intermediate scattering function (SISF):
\begin{equation}
F_S(Q,t)= \left< \sum_{i=1}^{N}
e^{i{\mathbf Q}\cdot [
{\mathbf r}_i(t)- {\mathbf r}_i(0)]} \right> \label{isf}
\end{equation}
which can be measured in an incoherent quasielastic neutron
scattering experiment.

The functions (\ref{grts}) and (\ref{isf}) contain the information 
which concerns the single particle dynamics
and can be directly evaluated from MD trajectories. Even if they are equivalent
while (\ref{isf}) would allow a more direct comparison
with experiments, when available, the function (\ref{grts}) gives
a more detailed and intuitive description of the motion of the
particle in the fluid.

We now discuss the behavior 
of VHSCF for water in its bulk 
supercooled phase for a comparison
with the confined case. 
We conducted MD simulations in the NVE ensemble with 216 water molecules.
The model potential used is the SPC/E. This potential,
at variance with ST2~\cite{geiger-st2}, shows a glassy MCT
behavior upon supercooling~\cite{gallo-prl}.
We cooled the system along the
$0$ MPa isobar characterized by a $T_C$ estimated to be $194$~K~\cite{starr}.
In order to equilibrate the system at such a supercooled 
temperature a run of $100$ ns was performed
with a timestep of $0.2$ fs.

In Fig.~\ref{fig:1} we show $4 \pi r^2 G_S (r,t)$ of the oxygens
as function of $r$ at different times.
We can clearly 
distinguish different time regimes mastered by the caging 
occurring in supercooled liquids. Looking at the SISF
in the inset of the same figure we see that for short times
($t < 0.04$~ps)  the system is in the ballistic region,
the corresponding $4\pi r^2 G_S (r,t)$ shows a sharp peak and decays rapidly
to zero. For intermediate times ($ 0.2 <t< 100 $~ps) the system is
in the $\beta$ relaxation region, well described by the MCT, and the ``cage
effect'', evidenced by the clustering of $4 \pi r^2  G_S (r,t)$,
is rather marked as we are right on the estimated MCT
crossover temperature. This temporal region corresponds to
the plateau of the SISF.
Finally in late the $\alpha$-relaxation region,
where the SISF is described by a stretched exponential decay, 
the system is entering the diffusive regime and long tails
develop in the $ 4 \pi r^2G_S (r,t)$ which extends to longer distances  
at increasing time.
This behaviour of the VHSCF of supercooled water is
shared by most liquids approaching the MCT crossover temperature 
$T_C$.

\begin{figure}
\includegraphics[width=8cm]{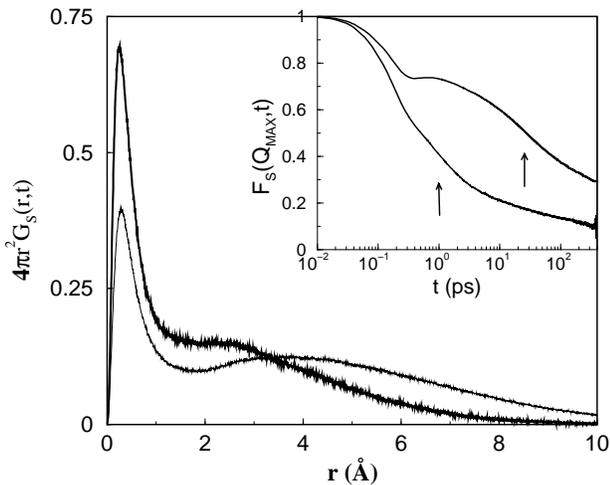}
\caption{In the main frame the total VHSCF for the oxygens of the
water molecules for T=298 K and t=1ps (lower curve) and T=210 K and t=25.6 ps.
In the inset the corresponding total SISF (lower T on the top).
The arrows mark the times in the $\alpha$-relaxation region 
at which the VHSCF are evaluated.}
\protect\label{fig:4}
\end{figure} 

\begin{figure}
\includegraphics[width=8cm]{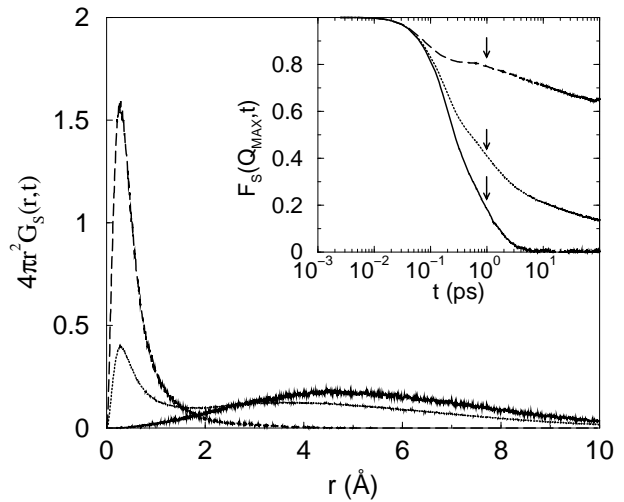}
\caption{In the main frame layer analysis of the VHSCF
for the oxygens of the water molecules at T=298 K and t=1ps. 
The total correlator
(dashed line) also displayed in Fig.\protect\ref{fig:1}, is shown together
with the contribution coming only from oxygens that move in the
inner part of the pore (continuous line) i.e. of the free water
and with the contribution coming from the first two layers
close to the substrate (long dashed line) i.e. of the bound water.  
In the inset the corresponding layer analysis for the SISF (lines as in the
main frame). The arrows mark the time  
at which the VHSCF are evaluated.}
\protect\label{fig:5}
\end{figure} 

\begin{figure}
\includegraphics[width=8cm]{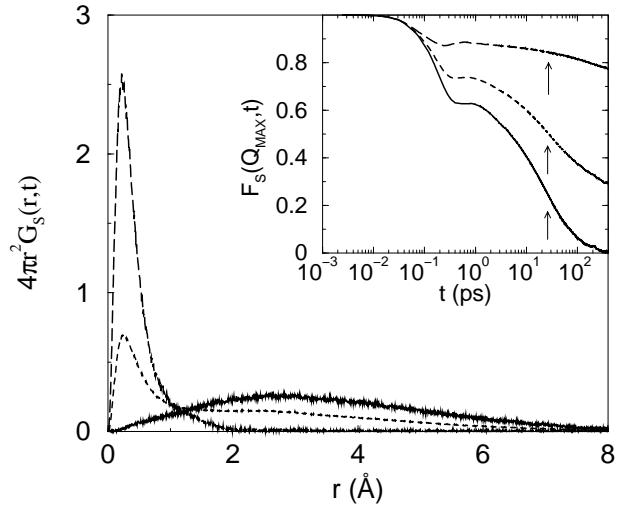}
\caption{In the main frame layer analysis of the VHSCF
for the oxygens of the water molecules at T=210 K and t=25.6 ps.
In the inset the corresponding layer analysis for the SISF.
The arrows indicate the times at which the VHSCF is evaluated.
}
\protect\label{fig:6}
\end{figure} 

\section{Layering effects in confined water}

We describe in this section our confined system and 
briefly summarize the main results
of the single particle analysis 
conducted in the Q,t space~\cite{noi-prl,noi-jcp,euro}.
We performed MD simulations  of SPC/E water in a
cylindrical pore of $40$~\AA\ diameter and $71.29$~\AA\ height. 
The pore is carved in a simulated silica glass. As described
in details in previous work~\cite{jmoliq} the surface of the pore is treated
in order to reproduce the main average properties of the
pores of Vycor glass. 
The substrate results to be a corrugated surface
composed by silicon atoms, bridging oxygens bound to silicons and
non-bridging oxygens saturated with acidic hydrogens. Different Coulomb
charges are attributed to the sites of the surface representing the 
four types of atoms, moreover the oxygen sites interact with the
oxygen sites of the water by means of a Lennard-Jones potential.
We analyze in the following the full hydration case which
corresponds to $2600$ molecules.
The pore surface results to be strongly hydrophilic, as can be seen from the
density profile shown in Fig.~\ref{fig:3}. 
A double layer of water molecules is clearly seen close to the
surface of the Vycor in the region between $15$ and $20$~\AA, 
where the pore surface is located. 
Moving toward the center of the pore
the density reaches a value close to
$11 \%$ of the bulk density at room temperature as observed
in the experiments on water confined in Vycor at full hydration~\cite{mar12}.
The figure shows that the effect of the temperature on the density profile 
is very small. The molecules
residing in the double layer structure 
display a strong 
distortion of the HB network~\cite{jmoliq,grr,rap}.

By analyzing the self intermediate scattering function 
we found that, due to the presence of strong inhomogeneities in our system,
a fit of the $\alpha$-relaxation region of the 
correlators to an analytic shape could be carried
out only by excluding the subset of molecules in the double layer close to
the substrate ($R>15$~\AA), identified with the so called bound water.
The subset of molecules belonging to the bound water appears to be in a glassy
state with very low mobility already at ambient temperature.
The remaining water molecules show a dynamical 
behavior typical of a glass forming liquid upon supercooling.
In particular the free water inside the pore shows 
a diversification of relaxation 
times as supercooling progresses, similar to SPC/E
bulk water~\cite{gallo-prl}. 
Focusing on the SISF we found that
the intermediate time region develops the MCT predicted 
plateau that stretches upon supercooling.
The $\alpha$-relaxation region shows a stretched exponential decay.
From the relaxation times extracted from the fit to this 
function it is possible to estimate the crossover 
temperature for free water~\cite{noi-prl}.
Evidence of two distinct dynamical 
behaviors has been found in experiments on other 
confined fluids~\cite{melni,barut,kremer,Mckenna}.

\begin{figure}
\includegraphics[width=8cm]{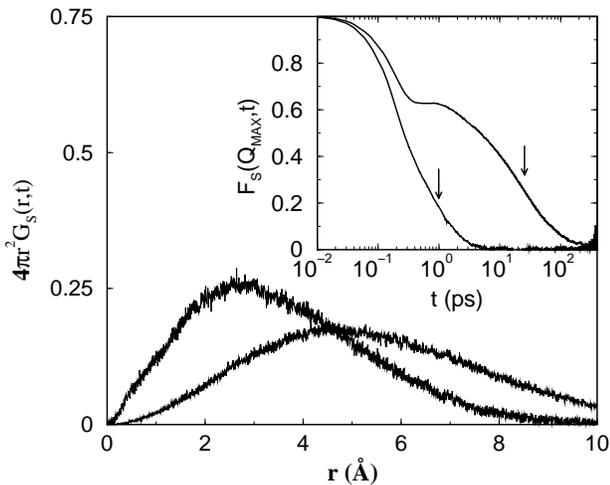}
\caption{
In the main frame VHSCF of free water for T=298K at t=1ps
(curve on the right) and T=210K at t=25.6 ps. In the inset the
corresponding SISF (lower temperature on the top).
The arrows indicate the times at which the VHSCF is evaluated.
}
\protect\label{fig:7}
\end{figure} 

\section{The Van Hove self correlation function of confined water}

We now move to the description of the dynamics of confined water
upon supercooling through the analysis of the VHSCF.

In Fig.~\ref{fig:4} we show the angular average of 
the VHSCF as defined in Eq.~(\ref{grts}), 
$4 \pi r^2 G_S (r,t)$, of the oxygens of the water molecules confined
in Vycor at $T=298$~K and $T=210$~K. 
For the sake of clarity only the highest and the lowest
temperatures investigated are shown in the pictures since
the intermediate curves display a continuous trend.
The curves
have been calculated for times corresponding 
to the late $\alpha$ region as best shown in the inset
where the corresponding SISF are plotted.
The more striking feature of the VHSCF is the presence 
in both curves of a double peak structure. 
A first sharper peak, located at short distance, is followed 
by a second broader peak. 

By considering the radial 
density profile of Fig.~\ref{fig:3} analogously to what done
for the SISF in previous work~\cite{noi-prl,noi-jcp,euro} we 
separated also for the VHSCF the contribution
of the molecules in the outer double layer close to the surface of
the cylindrical pore between $15$~\AA\ and $20$~\AA\ from the contribution
of the remaining molecules between the center of the pore and 
$15$~\AA.
The calculation is explicitly 
performed for a selected region
by considering as contribution to Eq.~(\ref{grts}) only
water molecules that at time $t=0$ are inside the considered region
until they remain in that region. 

In Fig.~\ref{fig:5} we report the inner layer (free water) contribution
and the outer layer (bound water) contribution to the total
VHSCF of Fig.~\ref{fig:4} at $T=298$~K.
In the inset we show
the same layer analysis performed for the intermediate scattering
function (\ref{isf}) of the confined water.
We note that the two VHSCF are singly peaked 
in the $\alpha$-relaxation region and that each peak corresponds
to one of the two peaks of the total VHSCF.
The VHSCF of the bound water is very localized and   
decays to zero in few~\AA, while in the free water 
the molecules are more distributed in space and the corresponding
VHSCF decays slowly.

In Fig.~\ref{fig:6} the same analysis is shown for T=210 K. Features
are similar to the previous figure. 
Interestingly, this division in two subset reveals 
to be temperature independent in our system 
and therefore a general characteristic
in the range of temperatures investigated for this time region. 

The VHSCF at $210$~K for $25.6$~ps is more localized close to the
origin with respect to the corresponding correlator calculated at
room temperature for $1$~ps, as shown in Fig.~\ref{fig:7}. 
This is
due to the slowing down of the dynamics as the
temperature is decreased to the region of the supercooled liquid.

\begin{figure}
\includegraphics[width=8cm]{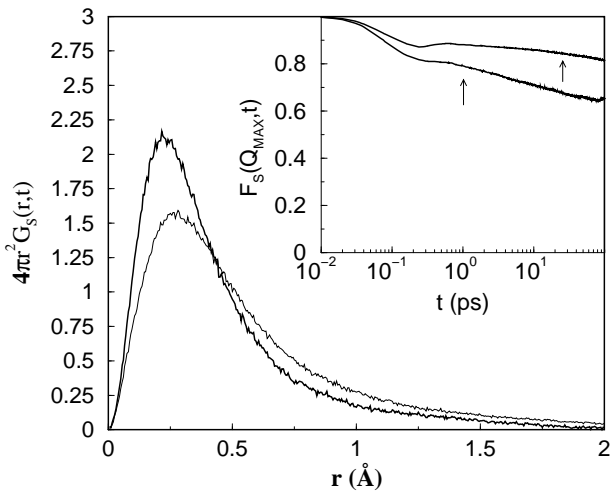}
\caption{In the main frame VHSCF of bound water for T=298K at t=1ps
(curve on the right) and T=210K at t=25.6 ps. In the inset the
corresponding SISF (lower temperature on the top).
The arrows indicate the times at which the VHSCF is evaluated.
}
\protect\label{fig:8}
\end{figure}

The contribution due to the molecules which reside closer to
the substrate (bound water) is reported separately in Fig.~\ref{fig:8}.  
It is evident that, at variance with
the free water, the bound water suffers
a severe slowing down of the dynamics
already at room temperature. The two distributions of
distances corresponding to T=298K and
T=210 K appear in fact not to differ substantially from each other. 

\begin{figure}
\includegraphics[width=8cm]{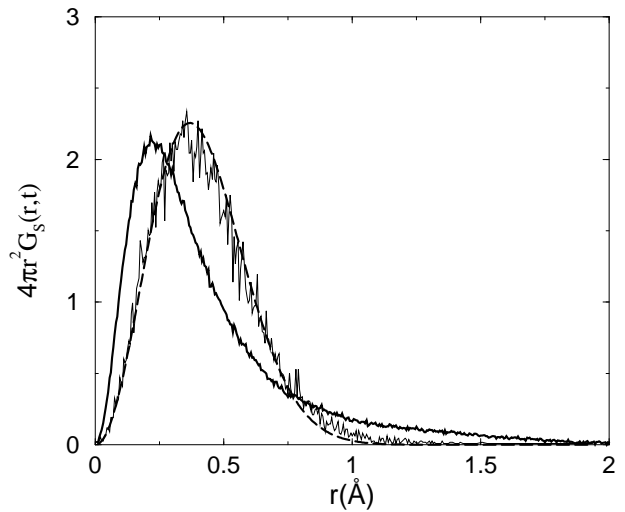}
\caption{VHSCF of bound water for T=210K at t=25.6 ps
(thick line) compared with VHSCF of bulk water for T=190K
at t=1 ps (thin line). A fit to a Gaussian is superposed 
to the VHSCF of the bulk (long dashed thick line).}
\protect\label{fig:9}
\end{figure}

The comparison of the VHSCF of supercooled bound water 
calculated in the $\alpha$ region with the VHSCF of bulk water
(see Fig.~\ref{fig:1}) shows that the
molecules are much less mobile in confined bound water. 
This is best seen in Fig.\ref{fig:9} where we report the VHSCF 
of confined water at T=210 K in the $\alpha$-relaxation region
with the most similar VHSCF of bulk water at T=190 K,
namely the one in the intermediate region of the rattling in the cage.
We note that in spite of the fact that the curves are 
calculated in two different dynamical regimes they
are very similar. Therefore the confined VHSCF reproduces
the features of molecules that can move essentially in the cage
of the nearest neighbors. 
The peak of the VHSCF curve 
of confined bound water is shifted left, due to
higher density of bound water with respect to the bulk.
It also displays a much longer tail.

\begin{figure}
\includegraphics[width=8cm]{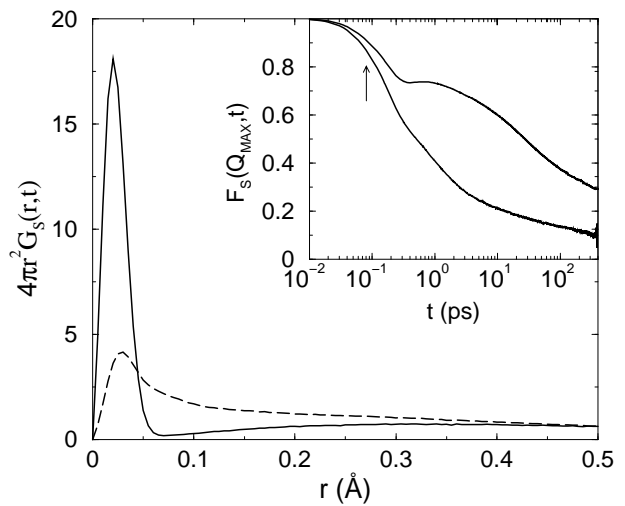}
\caption{In the main frame total VHSCF for T=298 K 
(dashed line) and T=210K (continuous line) both 
at t=0.08 ps. In the inset the
corresponding SISF (lower temperature on the top).
The arrow marks the time at which the VHSCF are evaluated.
}
\protect\label{fig:10}
\end{figure}

In Fig. \ref{fig:10} we report the total correlator calculated for
the highest and the lowest temperature investigated
in the early times. At a first glance
the behavior might seem quite similar to the late $\alpha$ one.
Nonetheless it turned out
that the shell analysis is not valid in this region.
We therefore conclude that our layer analysis is limited to the
temporal range of the $\alpha$-relaxation for this model 
while for early times the two dynamical regimes cannot
be distinguished.


\section{Summary and Conclusions}

We presented a study of the single particle Van Hove
correlation function for a model of water confined in a silica
pore of Vycor glass. 

The separation in two well distinct subsets of water molecules
in the hydrophilic pore hypotized in a previous analysis is
confirmed as unique. The VHSCF is in fact doubly peaked at all temperatures
and a layer analysis of the correlators shows 
that water close to the substrate, bound water, contributes only to  the
first peak while free water contributes only to the second peak. 
This separation is however valid only in the slow relaxation region
commonly named as $\alpha$. 
No separation of dynamical regimes has been found 
in the fast relaxation region.

Due to the strong resemblance that we found in bound water with respect
to water confined in biological environments~\cite{rap} we infer that the
double dynamical behavior of this system is 
shared by water confined in biological environments like surrounding proteins.

\end{document}